\documentclass[12pt]{iopart}
\pdfoutput=1
\usepackage{bmpsize}
\usepackage{graphicx,color}								
\usepackage{amssymb}
\usepackage{array}
\newcommand{\Tef}{T_{\rm eff}}

\begin{document}

\title{Controlling local order of athermal self-propelled particles}

\author{Niamh Dougan$^2$, Peter Crowther$^1$, C. Patrick Royall $^1$$^,$$^2$$^,$$^3$$^,$$^4$ and Francesco Turci$^1$}
\address{$^1$H.H. Wills Physics Laboratory, Tyndall Avenue, Bristol, BS8 1TL, UK}
\address{$^2$School of Chemistry, University of Bristol, Cantock's Close, Bristol, BS8 1TS, UK}
\address{$^3$Centre for Nanoscience and Quantum Information, Tyndall Avenue, Bristol, BS8 1FD, UK}
\address{$^4$Department of Chemical Engineering, Kyoto University, Kyoto 615-8510, Japan}

\ead{f.turci@bristol.ac.uk}

\begin{abstract}
We consider a model of self-propelled dynamics for athermal active particles, where the non-equilibrium active forces are modelled by a Ornstein-Uhlenbeck process. In the limit of no-driving force, the model reduces to the passive, Brownian dynamics of an atomistic glass forming fluid, the Wahnstr\"om binary mixture. The Wahnstr\"om mixture is known to show strong correlations between the emergence of slow dynamics and the formation of locally favoured structures based on icosahedra. Here, we study how the non-equilibrium driving affects the local structure of the system, and find that it strongly promotes icosahedral order. The states rich in local icosahedral order correspond to configurations of very low potential energy, suggesting that the non-equilibrium dynamics in the self propelled model can be effectively exploited to explore the potential energy surface of the binary mixture and have access to states that are difficult to attain using passive dynamics.
\end{abstract}

\section{Introduction}

Suspensions of active particles in a fluid, also termed \textit{active fluids}, are radically different from classical liquids. Bacteria \cite{Wu:2000ki,Sokolov:2009dy,Lopez:2015cv} and chemically driven Janus particles\cite{Golestanian:2007hu,Jiang:2010el, Buttinoni:2013de} are examples of systems which consume energy from the surrounding environment in order to perform a form of \textit{self-propelled} motion. This motion indicates that the systems are inherrently out of equilibrium and can lead to spontaneous phase separation \cite{Peruani:2006gl,Redner:2013jo,Fily:2014cp}, giant density fluctuations \cite{Zhang:2010jn,Fily:2012hj} and spectacular rheological behaviour \cite{Giomi:2010dsa}.

However, it is important to understand which existing concepts from the equilibrium thermodynamics of fluids (such as pressure \cite{Solon:2015hza,Speck:2016fc}, effective temperatures \cite{Preisler:2016ci}, or density fluctuations \cite{Henkes:2011ed}) can be used to characterise this new class of systems. In particular, when the density of active particles increases, a regime of slow dynamics is reached. This regime has been shown to share some similarities with the glassy dynamics of supercooled liquids \cite{Berthier:2013bf}: specifically, a rapid increase in relaxation times has been observed together with the appearance of collective motion \cite{Angelini:2011dm,Berthier:2014ej}.

A more subtle feature of several models of passive glass-forming fluids is the growth of structural correlation lengths when the system is cooled \cite{coslovich2007, malins2013jcp, royall2015physrep}. Some models, especially, show that the emergence of slow dynamics is strongly correlated with the formation of long-lived, localised structures, corresponding to local minima in the potential energy landscape of the system\cite{hocky2014}. Whether such minima play any role in the dynamics of dense active fluids is yet far from being understood \cite{royall2015physrep}.

In this work we consider the interplay of self-propulsion forces and local structure in a model of active fluids inspired by the passive dynamics of supercooled liquids \cite{Szamel:2015im,flenner2016nonequilibrium}. We consider on one hand the passive case as a reference system, whose local structure is fully determined, and then gradually increase the contribution of the non-equilibrium, self-propelling forces to the global dynamics. We then measure the dynamical and structural properties of the passive and active systems, revealing that the active dynamics lead the system to explore states with slower dynamics, richer in local structures and lower in potential energy. This not only supports the idea that active dynamics and glassy dynamics can coexist, but also suggests that non-equilibrium active forces can be used in order to facilitate the exploration of the potential energy landscape of glassy systems.

This article is organised as follows: in section \ref{sec:meth}, we introduce the model and the numerical methods used to analyse it; in section \ref{sec:stdyn} we report our results concerning the structural and dynamical features of the considered active model; we conclude with a discussion of the results and further perspectives.

\section{Model and methods}
\label{sec:meth}
\subsection{Numerical simulations}
We consider an equimolar additive binary mixture of Lennard-Jones large (A) and small (B) particles of diameters $\sigma_A=1.2\sigma_B$, with masses $m_A=2m_B$ at constant number density $\rho=N/V=1.296$, originally introduced by Wahnstr\"om \cite{wahnstrom1991}. We focus on large systems of $N=8000$ particles. The units used throughout this work are reduced Lennard-Jones units, with respect to the A particles. 

When coupled to a thermostat, the Wahnstr\"om model is a fragile glass former which shows a rapid increase in relaxation time when cooled below the onset temperature \cite{kim2013} $T_o\sim0.70$ at which glassy behaviour and dynamical heterogeneities become observable. It has been shown that the slowdown at lower temperatures is accompanied by the growth of local order, in the form of extended icosahedral domains \cite{malins2013jcp,hocky2014, pinney2015recasting}.

In the present work, we choose to use the original interactions of the model in the context of driven dynamics, realising a simple model of self-propelled active matter. In particular, we aim to control the degree of local order not by the means of a passive thermostat coupled to the dynamics of the particles (passive dynamics) but with the use of active dynamics that mimics the motion of self-propelled particles. To do so, we explicitly refer to a recently developed stochastic model for the motion of athermal self-propelled particles \cite{Szamel:2015im}, originally studied in the framework of Mode Coupling Theory.

Under the model's equations of motion, the system performs an Ornstein-Uhlenbek process on the positions $\vec{r}_i$ and the forces $\vec{f}_i$:
\begin{eqnarray}
\dot{\vec{r}}_i=(\vec{F}_i+\vec{f}_i)/\xi_0\\
\vec{f}_i=-\vec{F}_i/\tau_p+\vec{\eta}_i
\label{eq:model}
\end{eqnarray}
 where $\xi_0$ represents the friction of particle $i$ with an effective medium, $\vec{F}_i$ the conservative forces resulting from particle-particle Lennard-Jones additive interactions, $\tau_p$ is the persistence time of self propulsion, and $\vec{\eta}_i$ is a delta-correlated Gaussian noise with zero mean and variance $\langle\vec{\eta}_i(t) \vec{\eta}_j(t')\rangle =2\Tef\xi_0/\tau_p^2 I\delta_{ij}\delta(t-t')$, where $I$ is the unit tensor. $\Tef$ represents the single particle effective temperature, as discussed in \cite{Szamel:2015im}, related to the diffusivity of a noninteracting particle evolving according to Eq.~\ref{eq:model}: for non-interacting particles, the Ornstein-Uhlenbeck process reduces to a persistence random walk with diffusion constant $D_0=\Tef/\xi_0$. At fixed density $\rho$, in the limit $\tau_p\rightarrow0$, the effective temperature corresponds to the temperature of a Brownian system $T=\Tef$. We consider two effective temperatures $\Tef=0.64,0.60$ below the onset temperature of the passive system $T_o$, for which dynamical heterogeneities and growing local order have been observed.

\subsection{Identifying local structure}
The passive Wahnstr\"om mixture presents the formation of local domains of icosahedral order at sufficiently low temperatures \cite{malins2013jcp,royall2014}. These are strongly correlated with the emergence of slow, heterogeneous dynamics \cite{hocky2014}, and have also been observed to induce crystallisation into a Frank-Kasper phase \cite{pedersen2010}.
 
Icosahedral local order is detected using the Topological Cluster Classification (TCC) algorithm \cite{malins2013tcc}. This relies on the identification of local neighbours with a Voronoi tessellation of the system. Only direct neighbours are retained in order to reduce the effect of thermal fluctuations and identify stable icosahedra uniquely. Icosahedra are particularly relevant for the Wahnstr\"om binary mixture as they correspond to the local minimum of the potential energy for isolated clusters of 13 particles \cite{malins2013jcp}.

Following the classification provided by the TCC, we also keep track of two additional local motifs that are not related to the icosahedral order: an 11-particle cluster (termed 11W) which also is a minimum energy cluster of 11 Wahsntr\"om particles at different ratios of A and B particles \cite{malins2013jcp} and the crystalline face centred cubic cluster (fcc) composed by 15 particles. Other clusters in the TCC classification were monitored, but they do not contribute significantly to the observed behaviour of the system and so we do not discuss them further.

\section{Dynamics and structure}
 \label{sec:stdyn}
 \subsection{Pair correlations}
 
\begin{figure}[t]
	\centering
	\includegraphics[scale=1.25]     {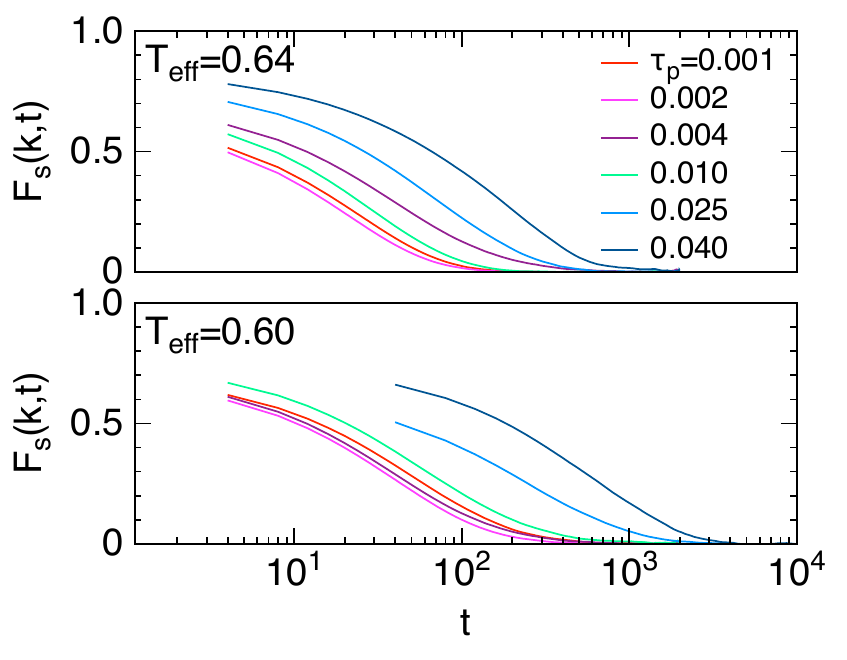}
	\caption{Self part of the intermediate scattering function $F(k,t)$ at $k=2\pi/\sigma_A$ for different persistence times $\tau_p$ at the two sampled effective temperatures $\Tef=0.64,0.60$.}
	\label{fig:isfs}
\end{figure}

\begin{figure}[t]
	\centering
	\includegraphics[scale=1.25]     {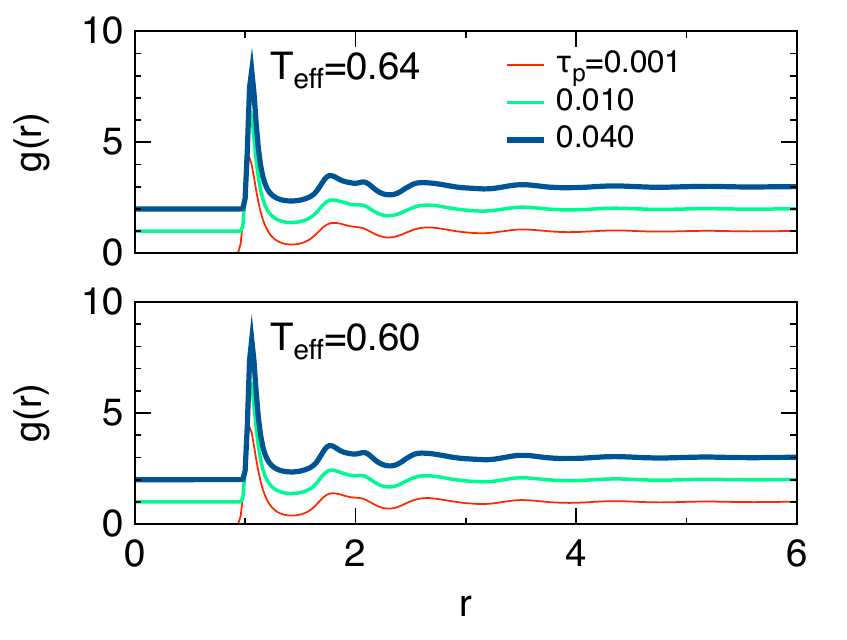}
	\caption{Radial distribution functions for two effective temperatures and three values of the persistence time $\tau_p$. Notice the splitting of the secondary peak. The curves are shifted upwards by a unit.}
	\label{fig:gr}
\end{figure}

We first characterise the active system in terms of pair correlations of dynamical and structural observables. The dynamics is probed by measuring the self part of the intermediate scattering function $F_s(k,t)$
\begin{equation}
	F_s(k,t)=\left\langle \sum_{i=1}^N e^{-i\vec{k}\cdot(\vec{r}_i(t)-\vec{r}_i(t_0))}\right\rangle,
\end{equation} 
at $k=2\pi/\sigma_A$ and averaged at steady state. For the structure at the pair level, we compute the radial distribution function:
\begin{equation}
g(r)=\frac{1}{N\rho}\left\langle\sum_i\sum_{j\neq i} \delta(r-|\vec{r}_i-\vec{r}_j|)\right\rangle,
\end{equation} 
where $N$ is the number of particles and $\rho$ is the number density. The intermediate scattering function (ISF) and the radial distribution function $g(r)$ of the active system are shown in Fig.~\ref{fig:isfs} and~\ref{fig:gr} respectively for two effective temperatures, $\Tef$, and a wide range of persistence times.

\begin{figure}[t]
	\centering
	\includegraphics[scale=1.25]     {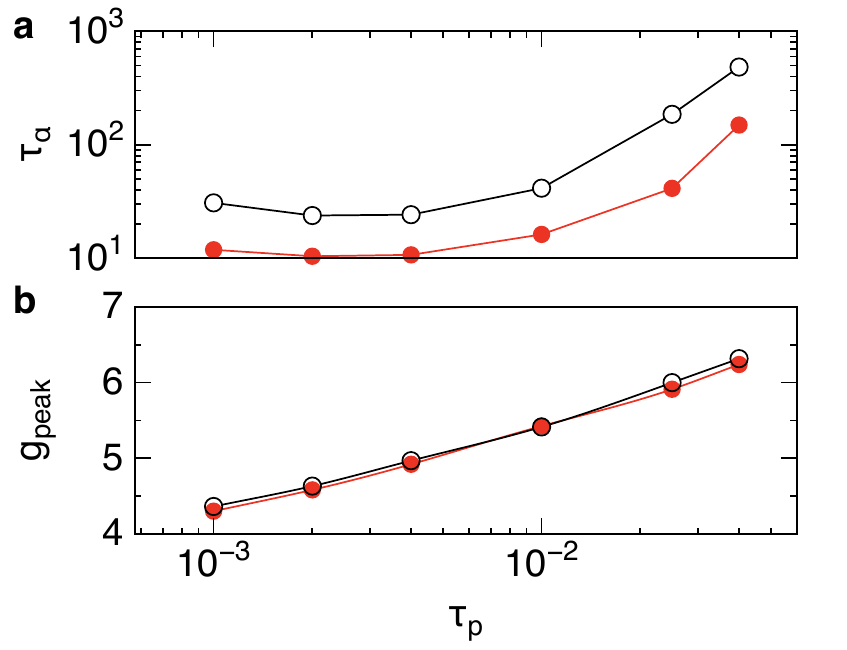}
	\caption{Relaxation times and local order. (a) Non monotonic behaviour of the relaxation times $\tau_{\alpha}$ as a function of the persistence time $\tau_p$ compared to the monotonic increase of the contact number as determined from the height of the first peak $g_{\rm peak}$ (b) in the radial distribution function g(r) at the effective temperature $\Tef=0.64$ (filled symbols) and $\Tef=0.60$ (open symbols).}
	\label{fig:taus}
\end{figure}

From the ISF we can determine the timescale for relaxation. By modelling the decay of the intermediate scattering function (Fig.~\ref{fig:isfs}) with a stretched exponential function $F_s(k,t)\sim\exp[-(t/\tau_\alpha)^\beta]$, we can extract the characteristic $\alpha$-relaxation time $\tau_{\alpha}$ of the system. In the passive case, this is normally interpreted as the time taken for a particle to escape from the cage formed by its nearest neighbours. We notice that at the two effective temperatures $\Tef$ considered, the relaxation time varies non-monotonically with increasing persistence time. As shown in Fig.~\ref{fig:taus}(a), the systems that appear to relax the fastest have an optimal value of persistence times $0.002<\tau_p^\ast<0.004$.

This non-monotonic behaviour has been observed in other active systems, notably in mono-disperse Lennard-Jones particles and the Kob-Andersen binary mixture \cite{Szamel:2015im}, which have local dominant order based on the fcc crystal and the bicapped square antiprism respectively \cite{coslovich2007,malins2013fara,crowther2015}. Our work brings further evidence that this effect does not depend on the precise kind of local order, as the Wahnstr\"om mixture's icosahedral local order is different from both the Kob-Andersen and Lennard-Jones fluid. This indicates that the non-monotonicity of the relaxation time as a function of the persistence time is a signature of the self-propelled dynamics.

Analogously to the Lennard-Jones and Kob-Andersen case, the non-monotonicity in the relaxation times is accompanied by a monotonic structural response as the persistence time is increased. In Fig.~\ref{fig:gr} we see that longer persistence times correspond to an increase of the amplitude of the first peak of the radial distribution function, and that further restructuring occurs at the level of the second and third coordination shells, with the formation of a shoulder which is often encountered in systems characterised by icosahedral motifs \cite{dzugutov1992}. As detected by the first peak of $g(r)$, the increase of order is very weak and proportional to $\log{\tau_p}$.
 
 We also notice that when the effective temperature $\Tef$ is reduced, the system becomes slightly slower, with marginally stronger structural correlations in the $g(r)$ (see Fig.~\ref{fig:taus}(b)) but the optimal value of the persistence time for which the relaxation is the fastest does not vary significantly.

\subsection{Local order}
We now consider the higher order structural changes that the system undergoes when varying the persistence time $\tau_p$ and the effective temperature $\Tef$. First, we demonstrate that even in the self-propelled case local icosahedral order forms the dominant, long-lived motifs that characterise the system.

\begin{figure}[t]
	\centering
	\includegraphics[scale=1.25]     {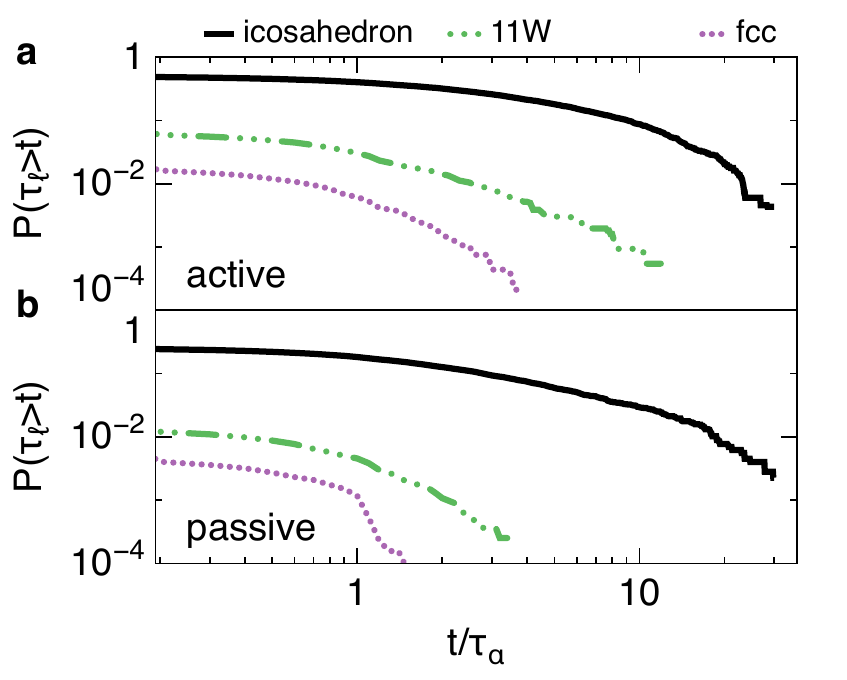}
	\caption{Cumulative probability distributions of lifetimes $\tau_\ell$ for the active (a) and passive (b) case at temperature $\Tef=0.60$ for several local clusters. The active case is measured at $\tau_p=0.04$. }
	\label{fig:lifetimes}
\end{figure}

To do so, we employ the TCC algorithm with several candidate structures that contribute to local order and compute their respective life times. The lifetime is defined as follows: we consider the particles as distinguishable and we label them with indices; we then identify a given instance of a structure of type $X$ at time $t$ as the set of indices of the particles forming it; if a subset of the particles in the structure becomes un-bonded from the others after a time $\tau_\ell$, we record $\tau_\ell$ as the life-time of the original instance. This allows us to produce cumulative probability distributions $P(\tau_\ell>t)$ for structures with lifetimes $\tau_\ell$ longer than $t$. Fig.~\ref{fig:lifetimes} shows that the overall trends for the passive case (corresponding to the $\tau_p=0$ limit) and the active case with longest persistence time $\tau_p=0.04$ are similar. Notice that, among all the considered structures, the icosahedron always has the longest tails in its cumulative probability distribution of lifetimes.

Focusing on the three selected local structures that we selected (the icosahedron, 11W and fcc), we notice that while the cumulative probability distributions related to icosahedral order are similar in the active and passive case, the cumulative probability distributions for the fcc and the 11W have longer tails in the active system. This indicates that the effect of longer persistence times on different types of local structure is to extend their lifetimes, correlating the motion of the particles and delaying the dissolution of locally ordered motifs.

Analogously to the first peak of the pair correlation function, the degree of icosahedral order is a monotonically increasing function of the persistence time $\tau_p$. This is quantified by the average concentration of particles detected in icosahedral domains $n=N_{\rm ico}/N$ at steady state, as shown in Fig.~\ref{fig:nico}(a). We notice that for both the two effective temperatures studied, long persistence times lead to a striking increase in the degree of icosahedral order, with more than double the concentration of icosahedra than in the corresponding passive case. We also investigated persistence times beyond $\tau_p=0.04$ and the nucleation of long-range order was observed (similarly to that which occurs to Lennard-Jones particles in the same limit).

\subsection{Fractal dimension and potential energy}

In order to quantify how the shape of the icosahedral domains changes as a function of the persistence time, we estimate their fractal dimension. We compute the Hausdorff dimension\cite{Mandelbrot:1983tz} (typically smaller than the Euclidean dimension) following the box counting algorithm \cite{Gagnepain:1986ho}, which subdivides the system into cells of variable linear size $s$ and evaluates the number of cells $N_c$ occupied by a particle within an icosahedral domain as a function of $s$. The box-counting fractal dimension is defined as, 

\begin{equation}
	d_f=\lim_{s\rightarrow 0} \frac{\log N_c(s)}{\log 1/s}.
\end{equation}

\begin{figure}[t]
	\centering
	\includegraphics[scale=1.25]     {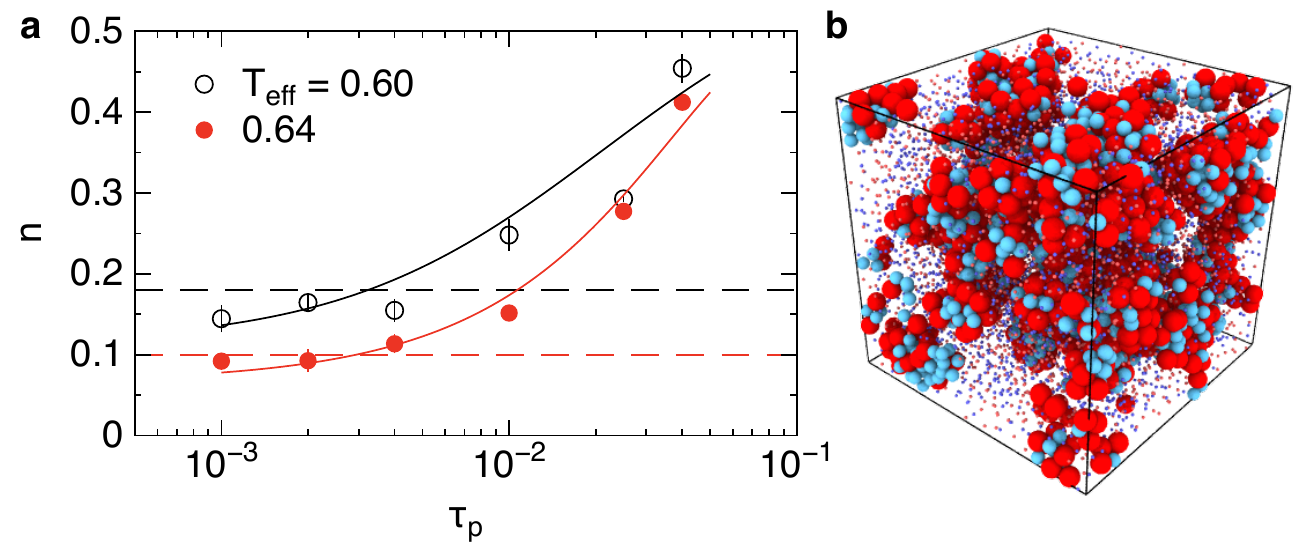}
	\caption{(a) Fraction of particles $n=N_{\rm ico}/N$ found in domains of icosahedral order at the two sampled effective temperatures $\Tef=0.60,0.64$ (open and filled symbols respectively). Horizontal dashed line represent the concentrations for the inactive classical glass former at the temperature $T=\Tef$. The lines through the points are guides to the eye. (b) Snapshot at $\tau_p=0.025$ and $\Tef=0.64$ representing A (red) and B (blue) particles in icosahedral domains (large spheres) and outside icosahedral domains (small dots). The corresponding fraction of icosahedral particles is $n=0.24$. }
	\label{fig:nico}
\end{figure}

\begin{figure}[bt]
	\centering
	\includegraphics[scale=1.25]     {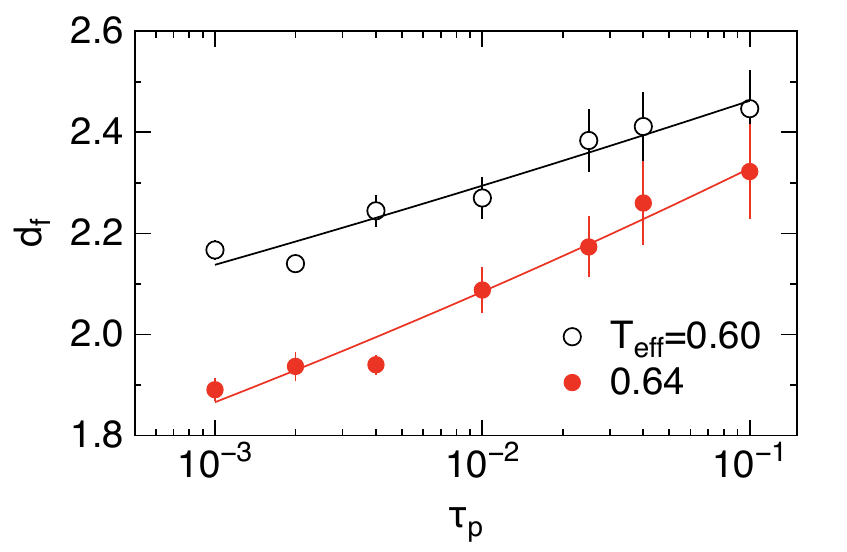}
	\caption{Fractal dimension from box counting $d_f$ of domains of icosahedral order at the two sampled effective temperatures $\Tef=0.60,0.64$ (open and filled symbols respectively). The lines through the points are guides to the eye. }
	\label{fig:df}
\end{figure}

As shown in Fig.~\ref{fig:df}, the fractal dimension increases as the persistence time increases, approximately following the increase in the number of particles in icosahedral domains $n$. In particular, the fractal dimension passes from $d_f \lesssim 2$ for $\tau_p\approx 10^{-3}$ to $d_f\approx 2.5$ for $\tau_p\approx 10^{-1}$, suggesting that branched icosahedral domains icosahedral domains (Fig.~\ref{fig:nico}(b)) become much more compact and space filling when the persistence time is increased.

\begin{figure}[bt]
	\centering
	\includegraphics[scale=1.25]     {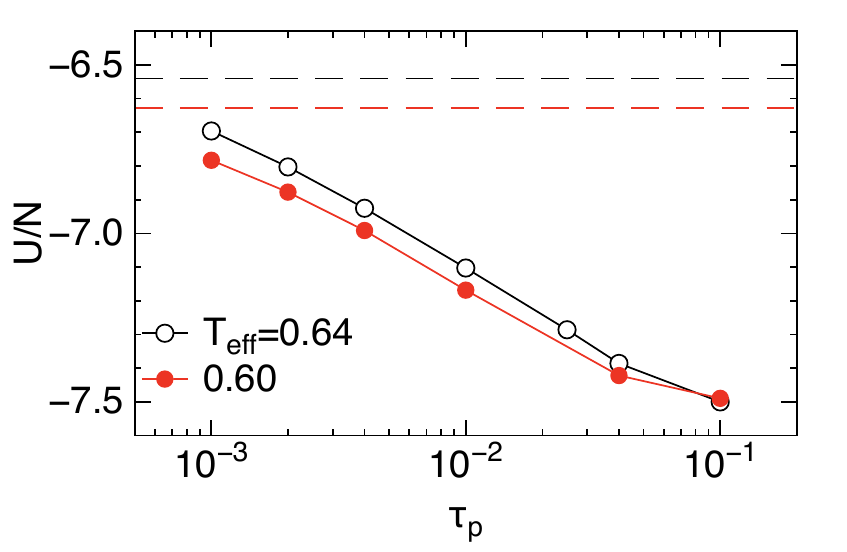}
	\caption{Potential energy per particle at steady state for the active system at effective temperature $\Tef=0.60,0.64$ (filled, empty circles respectively) for different persistence times $\tau_p$ compared to the passive system at the corresponding temperatures (dashed horizontal lines). The lines through the points are guides to the eye.}
	\label{fig:epot}
\end{figure}

The increase in the number of particles in icosahedral domains and the increase in the overall fractal dimension of icosahedral domains can be interpreted as the formation of more and more stable structures with fewer and fewer defects. The icosahedral order minimises the energy locally, and the overall increase in icosahedral order, as shown in Fig.~\ref{fig:epot}, leads to to a radical decrease in the potential energy per particle $U/N$. We notice that even moderately low persistence times induce a marked reduction in the potential energy per particle compared to the passive case. This indicates the formation of large low energy regions dominated by icosahedral order, compatible with the previously observed extended clusters of Frank-Kasper phases \cite{pedersen2010}.

\section{Conclusions}
We have considered a model of self-propelled dynamics based upon the Wahnstr\"om binary mixture. We tuned the strength of self-propulsion forces and compared the dynamical and structural properties of the system with the properties of the corresponding passive system. Both the passive and the active system show extensive regions of locally favoured structures, dominated by icosahedral order.

When the persistence time of the self-propulsion force is long, a strong structural signature is observed, with the formation of large icosahedral domains throughout the system. The active forces, appear to strengthen the relevance of locally favoured structures of icosahedral nature and allow us to access to very low energy regions of the potential energy landscape.

Differently from the passive case, the growth of spatial correlations in the active model is not a predictor of slow dynamics: there is an optimal value of the persistence time for which the relaxation is the fastest and the local order is only minimally affected. This signifies that for the active system there is no immediate mapping between the degree of local order and dynamical arrest.

The structural and dynamical changes can be interpreted as a consequence of increasing spatial correlations in the velocities \cite{Szamel:2015im,flenner2016nonequilibrium}, imposed by the longer persistence time in the dynamical equations. However, our discussion supports the idea that the glassy features of the model can be enhanced by the presence of active forces, and not reduced. This behaviour is different from what has been observed in hard sphere systems \cite{Stuart:2013kt,Berthier:2014eja}. In the studied mixture (which is in principle compatible with a Frank-Kasper phase \cite{pedersen2010}), we observe indeed the formation of an icosahedra-rich phase, that is amorphous in a wide range of parameters. This is related to a speedup of the dynamics for weak active forces (similarly to hard spheres) but to a slowdown for higher active forces, effectively allowing us to investigate very structural regions of the configurational space. In this sense, it is possible to consider the non-equilibrium self-propelled dynamics of Lennard-Jones fluids as a possible tool to explore regions of the potential energy landscape of glass-forming liquids which are normally hard to access, and eventually infer structural information on the low temperature regime of the passive models. This could be done, for example, considering the speedup in relaxation times introduced by weak active forces, which appear to only slightly perturb the structural feature of the mixture.

\section*{Acknowledgements}

CPR acknowledges the Royal Society for funding and Kyoto University SPIRITS fund. FT and CPR acknowledge the European Research Council (ERC consolidator grant NANOPRS, project number 617266).

\vspace{2cm}  

\bibliographystyle{iopart-num}
\bibliography{biblio.bib}
\end{document}